\documentclass[aps,preprint,superscriptaddress,showpacs, showkeys]{revtex4-1}

\usepackage{graphicx}
\usepackage[utf8x]{inputenc}
\graphicspath{{images/}}
\usepackage[a4paper,pdftex]{hyperref}
\usepackage{subfigure}
\hypersetup{
  pdftitle={Redimensioning the charge-density wave in TiSe2 by variation of the conduction band population},
  pdfborder={0 0 0 0}
}
\usepackage{color}

\begin{document}

\title{Realignment of the charge-density wave in TiSe$_2$ by variation of the conduction band population}

\author{Matthias M. May}
\email{Matthias.May@helmholtz-berlin.de}
\affiliation{Department of Physics, Humboldt-Universit\"at zu Berlin, Berlin, Germany}
\affiliation{Institute for Solar Fuels, Helmholtz-Zentrum Berlin f\"ur Materialien und Energie GmbH, Berlin, Germany}
\affiliation{Department of Chemistry, University of Cambridge, Cambridge, United Kingdom}
\author{Christoph Janowitz}
\affiliation{Department of Physics, Humboldt-Universit\"at zu Berlin, Berlin, Germany}
\author{Recardo Manzke}
\affiliation{Department of Physics, Humboldt-Universit\"at zu Berlin, Berlin, Germany}

\date{\today}

\begin{abstract}

We study the dependence of the dimensional character of the charge-density wave (CDW) in TiSe$_2$ by a controlled variation of the number of available charge carriers in the conduction band. A change of dimensionality is realised by H$_2$O exposure and monitored by angle-resolved photoelectron spectroscopy. The CDW spanning vector changes from 2x2x1 towards 1x1x2 in a narrow population range, just before a complete suppression of the CDW. Our findings enable a new view on the excitonic origin of the CDW, the relation between CDW spanning vector and periodic lattice distortion, and suggest modifications in the picture of the recently discovered chirality of the CDW.

\end{abstract}

\maketitle
\section{Introduction}

Charge-density waves (CDWs) have attracted much interest since their discovery in transition-metal dichalcogenides (TMDCs) almost four decades ago \cite{Wilson_CDW_TMDCs}. The question what drives these systems into a distorted state of high correlation has not only been a matter of discussion in condensed matter physics ever since, but has also engaged the non-linear dynamics community \cite{Strogatz_predicted_power_laws}. TiSe$_2$ is a particularly well-known representative of the TMDCs, exhibiting a CDW phase for which various driving forces have been proposed, such as an excitonic insulator transition \cite{Wilson_Excitonic}, a Jahn-Teller effect \cite{Anderson_Jahn-Teller}, electron correlation \cite{zhu_origin_2012} and exciton-phonon coupling \cite{Wezel_exciton_phonon, Monney_excitons_PLD}. The nature of its initial state as a semiconductor \cite{Rasch_PRL} or semi-metal \cite{Calandra_electronic_structure} is still under active discussion. As a model system for the CDW phase transition, TiSe$_2$ has recently awakened renewed interest upon the discovery of chirality \cite{ishioka_chiral_2010} and a superconducting phase, induced for instance by Cu-intercalation \cite{morosan_superconductivity_2006}, competing with the CDW phase. It was also argued that the chiral CDW actually requires a combination of excitonic insulator and electron-phonon coupling \cite{Zenker_Chiral_charge_order_TiSe2_2013}.

\begin{figure*}
  \includegraphics[width=\linewidth]{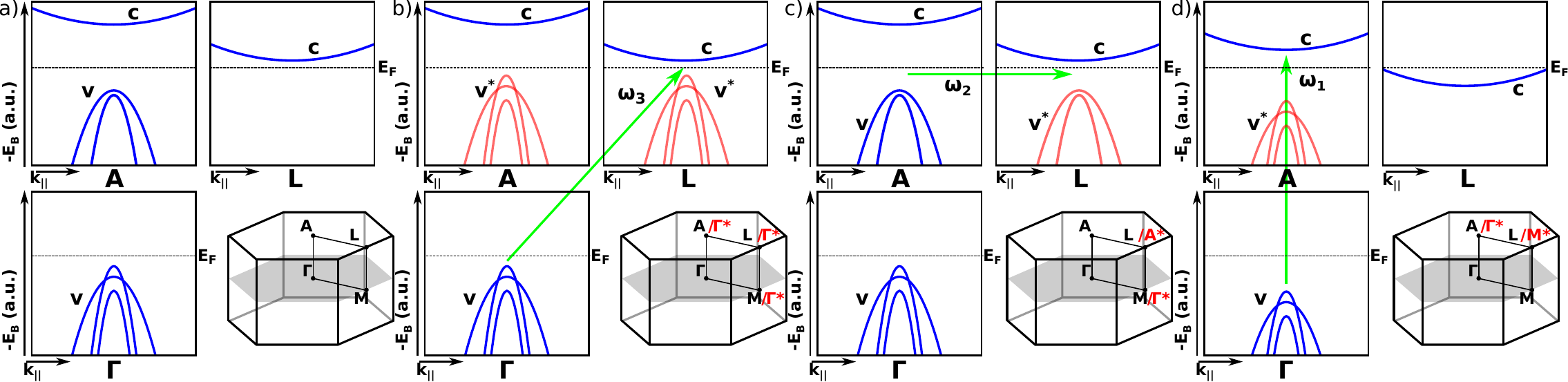} % figure 1: band scheme and BZ
  \caption{\label{fig: band scheme} Schematic band diagrams around the high-symmetric points $\mathit{\Gamma}$, \textit{A} and \textit{L} with corresponding Brillouin zones before the onset of the CDW phase (blue), adapted from \cite{zunger_band_1978, Cercellier_excitonic_insulator, Calandra_electronic_structure,Monney_temperature-dependent}. The conduction and valence bands are labelled \textit{c} and \textit{v}, respectively, the Fermi level $E_F$, the binding energy $E_B$ and the CDW spanning vectors $\vec{\omega_i}$. Features related to the CDW superstructure are coloured red and marked with a star. The different spanning vectors, $\vec{\omega_3}$, $\vec{\omega_2}$, and $\vec{\omega_1}$ define a 2x2x2, a 2x2x1, and a 1x1x2 superstructure, respectively. \textbf{(a)}, \textbf{(b)}, and \textbf{(c)} show the situation without H$_2$O exposure, \textbf{(a)} in the normal phase, \textbf{(b)} stating a 2x2x2 CDW, and \textbf{(c)} a 2x2x1 CDW. The situation for a pure 1x1x2 phase is depicted in \textbf{(d)}. Energy shifts and renormalisation of the dispersion due to CDW formation have been neglected.}
\end{figure*}

The layered TMDC $1T$-TiSe$_2$ exhibits a trigonal symmetry, and the slabs of Se-Ti-Se are only weakly stacked by van der Waals interaction in the out-of-plane \textit{c}-direction. In our view, it is a semiconductor with a very small, indirect band gap of $E_g\approx150\,\mbox{meV}$ \cite{Rasch_PRL} between its Se \textit{4p}-derived valence band at the high symmetric point, $\mathit{\Gamma}$, and the Ti \textit{3d}-derived conduction band at \textit{L} in its normal phase [Fig.~\ref{fig: band scheme}(a)]. In the context of CDW fluctuations \cite{Rossnagel_Rb_TiSe} at room temperature, band gap determination via the combination of ARPES and water adsorption \cite{Rasch_PRL} is a unique technique to determine the magnitude of $E_g$ in the non-CDW phase. The main signatures of the CDW phase are the gradual onset of a commensurate 2x2x2 periodical lattice distortion (PLD) below 202\,K \cite{Di_Salvo_CDW, Wilson_Excitonic, Brown_1980}, evidenced by neutron diffraction \cite{Di_Salvo_CDW}, and an electronic superstructure visible to angle-resolved photoelectron spectroscopy (ARPES), generally also stated as 2x2x2 \cite{zhu_origin_2012}. A prominent feature of the electronic superstructure is the appearance of backfolded valence bands at the symmetry points $L$ and $M$. This signature is accompanied by a significant transfer of spectral weight from the original valence bands to the backfolded bands, whose magnitude cannot be explained purely by the distorted lattice \cite{Cercellier_excitonic_insulator}. An experimental signature of the PLD, which is also typically used for the definition of the critical temperature, $T_c$, of the phase transition, is a peaking resistivity at $T_c$. The onset of this feature can be changed for instance by growth-related self-doping \cite{Di_Salvo_CDW, May_growth_conditions, Hildebrand_doping_nature_defects_TiSe_2014}, but the finding that a dampened resistivity peak does not necessarily coincide with a weakening of the backfolding evidenced by ARPES \cite{May_growth_conditions} hints towards a non-linear coupling of PLD and electronic superstructure.

A key question for understanding the phase transition in TiSe$_2$ is whether the appearance of the PLD is subordinated in the wake of an excitonic insulator instability or is an inherent feature of a second-order Jahn-Teller effect \cite{Rossnagel_Rb_TiSe}, dependent on strong electron-phonon coupling.

In this paper, we present experiments manipulating the conduction band population by H$_2$O adsorption, monitored by high-resolution ARPES. Our results question the generally stated 2x2x2 character of the CDW and establish the feasibility of a realignment from an electronic 2x2x1 reconstruction towards a 1x1x2 reconstruction. The interpretation of our findings in the context of existing models and the chirality of the CDW show that the electronic superstructure in the undisturbed CDW phase does initially not coincide with the 2x2x2 PLD, probably due to anisotropic electron-phonon coupling. Furthermore, they support the view of an excitonic driving force. 

\begin{figure}[h]
	\includegraphics[width=.5\linewidth]{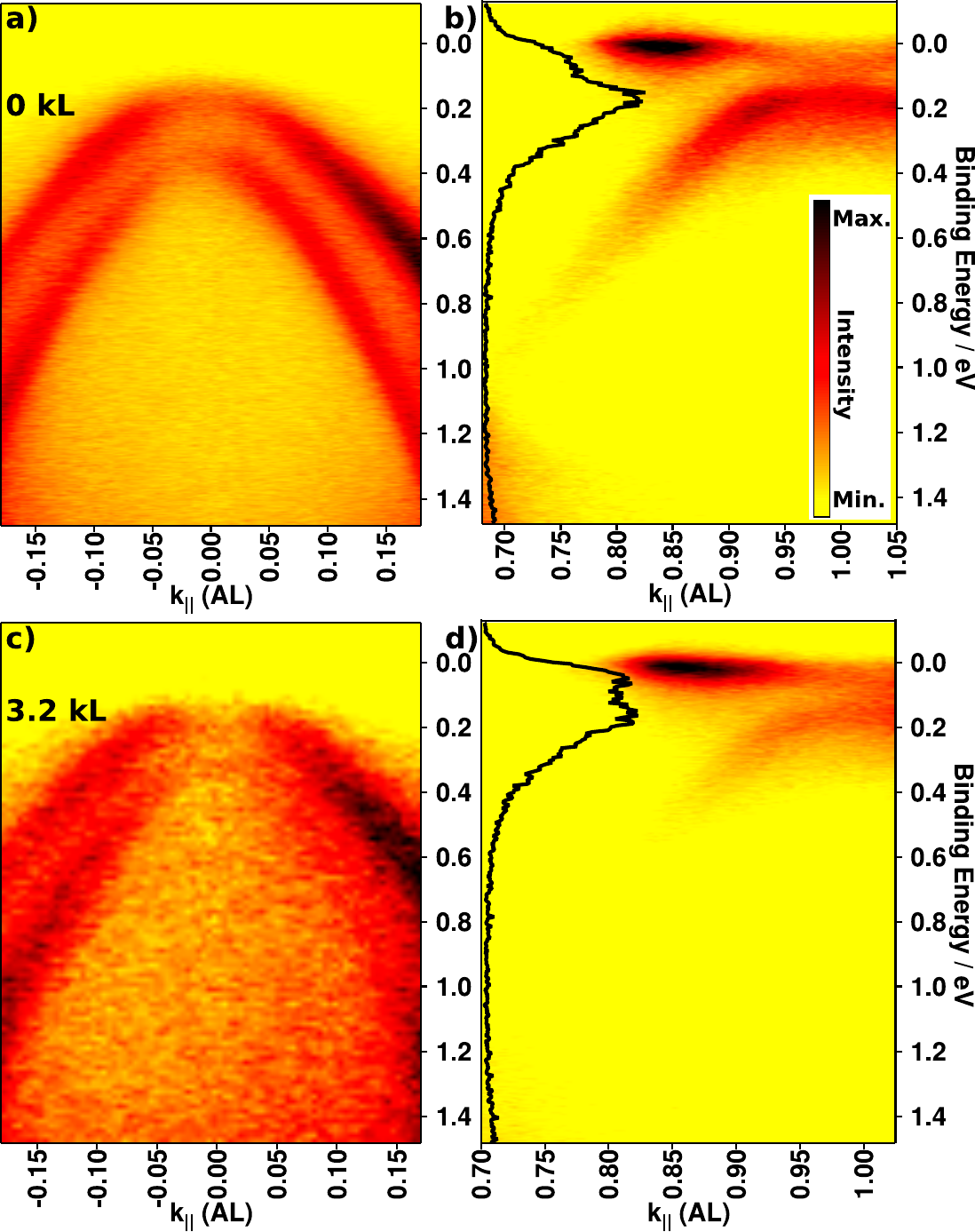} % figure 2
	\caption{\label{fig: AL no water} ARPES intensity maps around the high symmetric points \textit{A} \textbf{(a)} and \textit{L} \textbf{(b)} at T=21\,K without H$_2$O, and for \textit{A} \textbf{(c)} and \textit{L} \textbf{(d)} at T=22\,K with 3.2\,kL H$_2$O. Black curves in the spectra \textbf{(b)} and \textbf{(d)} show EDCs extracted exactly at the \textit{L}-point.}
\end{figure}

\section{Experimental}

The TiSe$_2$ single crystals studied here were grown by chemical vapour transport \cite{May_growth_conditions} using iodine as transport gas. Samples were cleaved in situ under UHV conditions. The ARPES experiments were carried out at the Beamline for Education and Scientific Training \cite{BEAST} at the synchrotron light source BESSY II with an energy resolution of 25\,meV and an angular resolution of $0.1\,^\circ$. Temperatures ranged from 20\,K to 300\,K and the photon energies of $h\nu=$22(31)\,eV, which probe the electronic structure around the high symmetric points $A$ and $L$ ($\mathit{\Gamma}$ and $M$), was used, scanning along the \textit{A-L} ($\mathit{\Gamma}-M$) direction. The water adsorption was inspired by previous experiments \cite{karschnick_adsorbate_1985,Rasch_PRL} demonstrating the reversible enhancement of the Ti \textit{3d} emission by physisorbed water. H$_2$O vapour was introduced into a dedicated UHV chamber on the cleaved samples at room temperature, resulting in a well-defined exposure on the surface, which was conserved upon transfer to the photoemission chamber. The adsorbed water creates a ``Schottky-like'' contact \cite{Rasch_PRL} inducing excess charge at the surface region, which is exactly the region probed by ARPES in the UV regime. Due to the layered structure of TiSe$_2$, free of surface states, the H$_2$O molecules are only physisorbed on the surface and a saturation of the population is found at an exposure corresponding to one monolayer of H$_2$O on the surface \cite{Rasch_PRL}. The mechanism is similar to electrostatic charge doping employed to induce phase transitions in MoS$_2$ by the application of an electric field via a gate electrode \cite{Biscaras_electrostatic_doping_MoS2_2015}.

\section{Results \& Discussion}

\begin{figure}[h]
	\includegraphics[width=.5\linewidth]{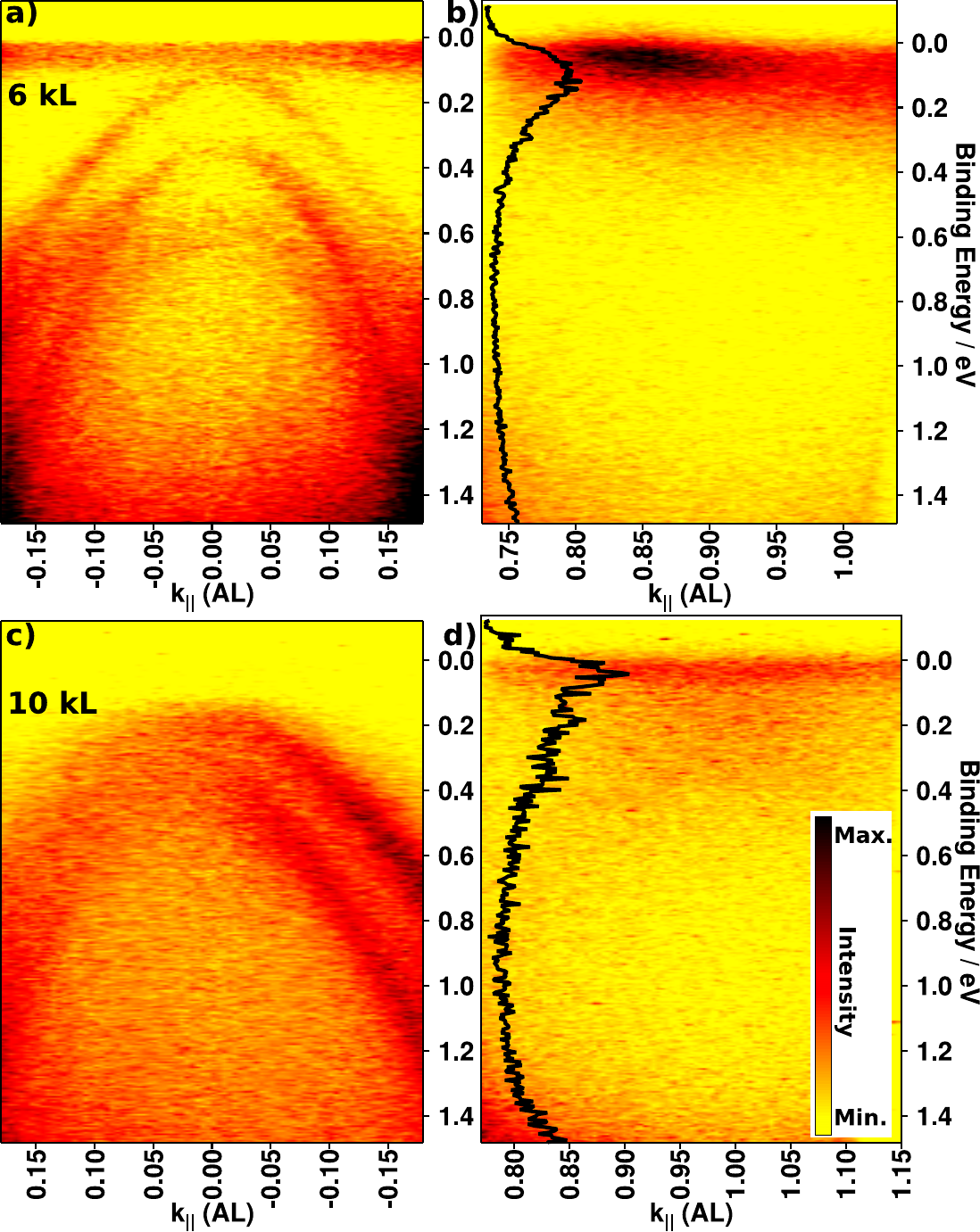} % figure 3
	\caption{\label{fig: AL 6kL} ARPES intensity maps around the high symmetric points \textit{A} \textbf{(a, c)} and \textit{L} \textbf{(b, d)} at T=41\,K with 6\,kL H$_2$O (a, b) and 10\,kL H$_2$O (\textbf{c, d}; 30\,K). The black curves in spectrum \textbf{(b) and (d)} show EDCs extracted exactly at the \textit{L}-point. The increase in noise with water exposure arises due to scattering of the photoelectrons by the adsorbate.}
\end{figure}

\begin{figure}[h]
	\includegraphics[width=.5\linewidth]{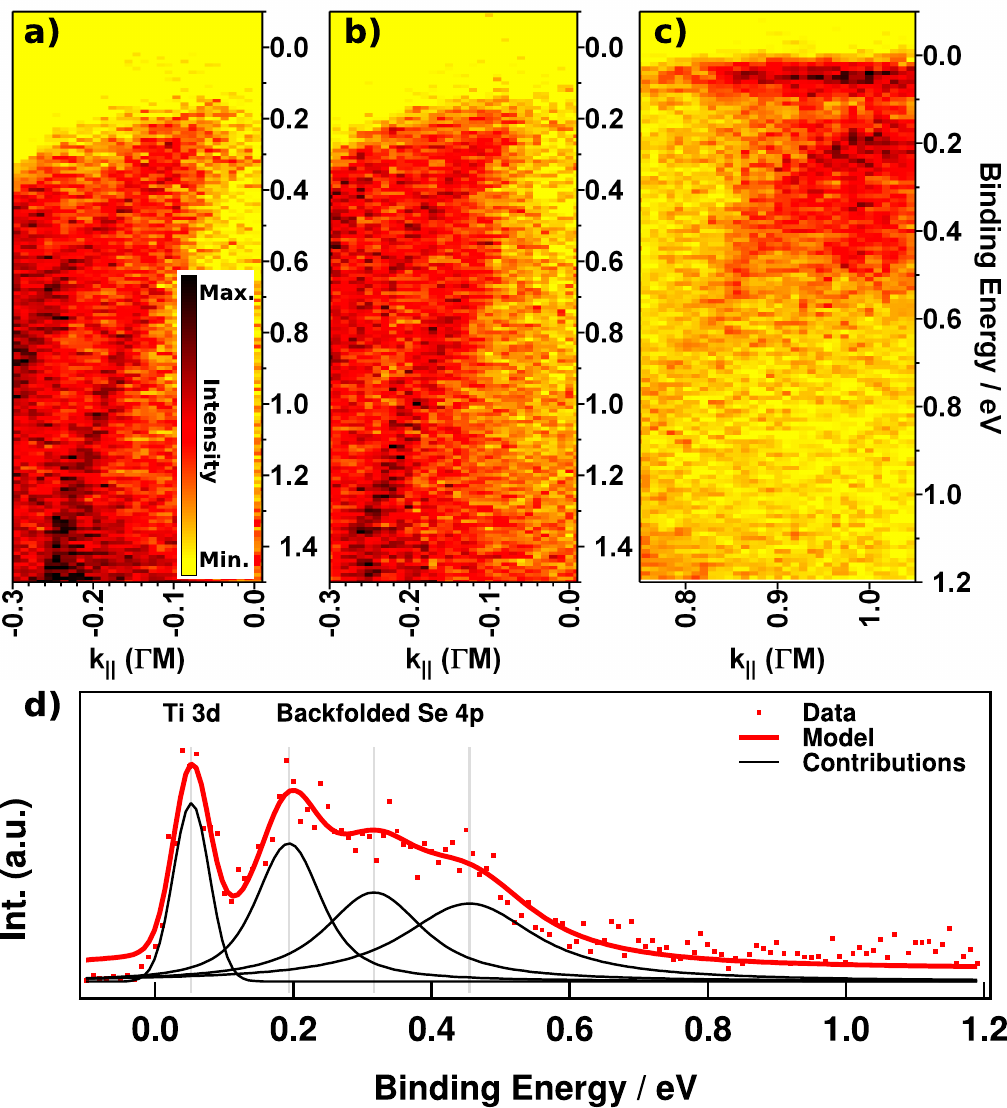} % figure 4
	\caption{\label{fig: Gamma_M} ARPES intensity maps around the high symmetric points $\mathit{\Gamma}$ \textbf{(a,b)} and \textit{M} \textbf{(c)} at T=20\,K. \textbf{(a,c)} are in the original, 2x2x1 phase prior to water exposure, while (b) is in the 1x1x2 phase. \textbf{(d)} shows an EDC extracted near the \textit{M}-point from \textbf{(c)}.}
\end{figure}

The CDW phase was monitored by ARPES, keeping track of its prominent experimental signature, the backfolded bands at high-symmetric points of the Brillouin Zone. The band structure at low temperature as observed by ARPES around $A$ [Fig.~\ref{fig: AL no water}(a)] in \textit{A-L} direction is characterised by two valence bands. At $L$, the backfolded valence band with hole-like dispersion is well-developed and parts of the Ti \textit{3d} conduction band can be identified near the Fermi level $E_F$ [Fig.~\ref{fig: AL no water}(b)]. Due to a coupling to the backfolded valence band, the shape of the conduction band emission is non-parabolic \cite{Kidd_electron_hole_coupling_TiSe_2002}. The origin of the backfolding shall be discussed in the following using a simplified band model (Fig.~\ref{fig: band scheme}) whose overall features were adapted from references \cite{zunger_band_1978, Cercellier_excitonic_insulator, Calandra_electronic_structure,Monney_temperature-dependent}. Calculated band structures \cite{zunger_band_1978, Calandra_electronic_structure} of the normal phase, indicated schematically in Figure~\ref{fig: band scheme}(a), suggest two valence bands at \textit{A} (in \textit{A-L} direction) and three at $\mathit{\Gamma}$ (in $\mathit{\Gamma}$\textit{-M} direction). The three bands observed experimentally at A and $\mathit{\Gamma}$ before and after water exposure are plotted in Figures~\ref{fig: AL 6kL}(a) and \ref{fig: Gamma_M}(a, b), respectively, which is consistent with the literature, where these bands were observed both in the normal as well as the CDW phase \cite{Cercellier_excitonic_insulator}. In a three-dimensional 2x2x2 CDW phase with a spanning vector $\vec{\omega_3}$, one would expect a $\mathit{\Gamma}^*$ point of the superstructure with three bands in $\mathit{\Gamma}^*$-\textit{M}$^*$ direction both at $A$ and $L$ [see BZ in Fig.~\ref{fig: band scheme}(b)]. Detailed analysis of energy distribution curves (EDCs), see also \cite{Monney_temperature-dependent, May_PRL}, at $L$ shows, however, only two peaks attributed to the backfolded valence band. At $A$, a decrease of spectral weight is observed, but no backfolding as one would expect for the new $\mathit{\Gamma}^*$. This rather suggests a 2x2x1 superstructure with a two-dimensional spanning vector $\vec{\omega_2}$, as depicted in Figure~\ref{fig: band scheme}(c), than a fully 3-dimensional 2x2x2 CDW. In this 2-dimensional phase, $\mathit{\Gamma}$ would be mapped to $M$ and $A$ to $L$, which is consistent with the observations: Near $M$, we do indeed observe three bands, as displayed in Figure~\ref{fig: Gamma_M}(c, d).

An exposure of the surface to H$_2$O and subsequent cooling down to low temperature reveals a gradual suppression of the 2x2x1 CDW phase: While the valence band at \textit{A} only changes slightly [Fig.~\ref{fig: AL no water}(c)], the backfolded valence band at \textit{L} [Fig.~\ref{fig: AL no water}(d), reproduced from ref.\cite{May_PRL}] shifts towards $E_F$ and its intensity decreases with increasing exposure at fixed temperature, just like the thermal melting of the CDW for the uncovered surface \cite{Monney_temperature-dependent}. This is in line with experiments which used in situ Rb-deposition \cite{Rossnagel_Rb_TiSe} and also found a gradual suppression of the CDW. An exposure of 3.2\,kL [Fig.~\ref{fig: AL no water}(c,d)] corresponds to a thermal melting equivalent to approximately 140\,K \cite{May_PRL}. Though an increase of water coverage shifts the valence bands away from $E_F$ (up to 130\,meV for the saturation coverage of one monolayer)\cite{Rasch_PRL}, the shift towards the Fermi level induced by the melting of the CDW is greater in magnitude, causing an overall shift of the valence band towards $E_F$ \cite{May_PRL}. Figure~\ref{fig: AL 6kL} shows the result of a relatively high exposure of 6\,kL. The backfolded band at \textit{L} largely disappears, while the Ti \textit{3d} emission becomes broader in $\vec{k}$- as well as energy-space [Fig.~\ref{fig: AL 6kL}(b)]. This indicates a further -- albeit not complete -- melting of the 2x2x1 CDW upon the high exposure level. Meanwhile, the situation at $A$ changes drastically. Instead of the two bands characteristic for the \textit{A-L} direction, three bands, as one would expect for $\mathit{\Gamma}$\textit{-M}, are reproducibly identified, see Figure~\ref{fig: AL 6kL}(a). These bands quickly disappear by increasing the temperature beyond 100\,K, which is significantly below the desorption temperature of the water molecules. Upon an exposure of 10\,kL, again only two bands are observed at low temperature, as can be seen in Figure~\ref{fig: AL 6kL}(c). This 1x1x2 phase, drafted in Figure~\ref{fig: band scheme}(d)\footnote{The backfolded valence band from $\mathit{\Gamma}$ and initial valence band at \textit{A} are energetically very close. Therefore, an interaction between the bands changing their dispersion can be expected, similar to that between initial and backfolded conduction bands \cite{Monney_temperature-dependent}. This will result in a different dispersion of the hole-like bands at \textit{A}}, exists therefore only in a narrow range of conduction band population and temperature and seems to imply a largely suppressed 2x2x1 phase [Fig.~\ref{fig: AL 6kL}(b)]. At 10\,kL H$_2$O coverage, it was still possible to resolve the split \textit{p}-derived bands at the \textit{A}-point [Fig.~\ref{fig: AL 6kL}(c)], while the split bands due to CDW influence had almost disappeared at the \textit{L}-point, i.e. the CDW had mostly vanished [Fig.~\ref{fig: AL 6kL}(d)] with the intrinsic band structure still intact.  A further exposure beyond 20\,kL suppresses the CDW completely \cite{May_PRL}.

Since our results indicate the possibility of the formation of a spanning vector in the third dimension (and its observability by ARPES), it is also an interesting question, why the 2x2x2 reconstruction [Fig.~\ref{fig: band scheme}(a)], clearly observed by neutron scattering \cite{Di_Salvo_CDW}, is not observable by ARPES in the initial electronic configuration. The difference may be based on the fact, that photoemission only monitors the density of occupied atomic orbitals, which do not necessarily coincide with the position of the cores, measured e.g. by neutron scattering. The apparent contradiction of a three-dimensional PLD alongside a two-dimensional CDW will be resolved in the following. Chirality of the CDW phase in TiSe$_2$ has been introduced recently \cite{ishioka_chiral_2010}. The underlying physics is that of a finite, anisotropic electron-phonon coupling described by van Wezel for TiSe$_2$ \cite{Wezel_chirality_and_orbital}: Generally, anisotropic electron-phonon coupling is characterised by the lattice distortion, $\vec{u}(\vec{x})$, which is related to the scalar modulation of the charge distribution, $\alpha(\vec{x})$, and the matrix elements $\eta_{ij}$ of electron-phonon coupling via $u_i\propto\sum_j\eta_{ij}\partial\alpha/\partial x_j$. Van Wezel \cite{Wezel_chirality_and_orbital} considers the case that the anisotropy of the \textit{p} orbitals in Te and Se leads to an anisotropic electron-phonon coupling for each orbital sector, resulting in a partly transversal polarisation of the ionic polarisation vector, $\vec{\epsilon}$. This vector, defining the overall direction of the PLD, will therefore in general not follow the CDW spanning vector, $\vec{\omega}$, longitudinally. Van Wezel states a purely transversal ionic displacement in TiSe$_2$, $\vec{\epsilon}\perp\vec{\omega_3}$ \cite{Wezel_chirality_and_orbital}, starting from a 2x2x2 CDW [Fig.~\ref{fig: band scheme}(a)] taking into account the anisotropy of the Ti and Se \textit{p} orbitals. Our finding of a 2x2x1 CDW [Fig.~\ref{fig: band scheme}(c)] however points to a slightly different situation. Since neutron scattering clearly reveals a 2x2x2 PLD \cite{Di_Salvo_CDW} and ARPES indicates an in-plane CDW with a two-dimensional propagation vector $\vec{\omega_2}$, the strict transversality $\vec{\epsilon}\perp\vec{\omega}$ cannot be realised, as $\vec{\omega_2}$ lacks an out-of-plane component. Instead, the ionic polarisation $\vec{\epsilon}$ is only partly transversal. Note that the lack of a component in the \textit{c}-direction can still enable the formation of a chiral phase \cite{Wezel_chirality_and_orbital}. As a consequence, a purely in-plane CDW does not conflict with a three-dimensional PLD or chirality.

The replacement of the electronic 2x2x1 in-plane superstructure by the out-of-plane 1x1x2 phase can be understood in the view of an exciton-driven CDW phase in the context of the band structure of TiSe$_2$. As soon as the water-induced population of the lowest-lying conduction bands at the \textit{L}- and \textit{M}-point reaches a threshold, excitons spanning \textit{AL} and $\mathit{\Gamma}$\textit{M} (i.e. the 2x2x1 vector) cannot be formed any more and this phase is largely suppressed. Thereafter, the unpopulated parts of the conduction band closest to the Fermi level can be found at $\mathit{\Gamma}$ and \textit{A}. The conduction band at \textit{A} is closer to $E_F$ than at $\mathit{\Gamma}$ due to its $\mathit{\Gamma}$\textit{-A} dispersion \cite{Calandra_electronic_structure} and therefore now energetically most favourable for the formation of excitons. This is also in line with the greater robustness of the CDW against an increase of the band gap than to a decrease \cite{May_PRL}. The number of charge carriers needed for this redimensioning lies presumably higher than the charge needed for the onset of a superconducting phase that could be invoked by additional carriers after by Cu-doping \cite{morosan_superconductivity_2006}. A hypothetical intermediate superconducting phase in the top layer(s) of TiSe$_2$, as already speculated upon by Rossnagel \cite{Rossnagel_Rb_TiSe}, induced by H$_2$O adsorption before the realignment of the CDW cannot be evidenced by the methods applied here.

Since the concept of a chiral CDW has been questioned recently \cite{Novello_STM_CDW_TiSe_defects_2015} it is instructive to check whether the presented results can be interpreted in favor of the chiral CDW. The 1x1x2 CDW [Fig.~\ref{fig: band scheme}(d)], which was described above, will most probably not exhibit chirality and the PLD should simply follow the CDW, $\vec{\omega_1}$. This could actually be related to the non-chiral CDW phase with a higher transition temperature, $T_{c,chiral}<T_{c,non-chiral}$, investigated experimentally as well as theoretically \cite{Castellan_Chiral_CDW_TiSe_2013, Zenker_Chiral_charge_order_TiSe2_2013}. The model of Zenker et al.\cite{Zenker_Chiral_charge_order_TiSe2_2013} does, however, not account for a change in dimensionality.

The transition temperature for the chiral CDW has been found to be only ca. 7\,K below the transition temperature to the ordinary CDW \cite{Castellan_Chiral_CDW_TiSe_2013}. It is most probable, that both temperatures shift to lower temperature upon H$_2$O uptake. It is remarkable that the 1x1x2 phase found here only exists in a narrow range of H$_2$O coverage which might be related to the narrow temperature interval between $T_{c,chiral}$ and $T_{c,non-chiral}$. In this narrow range, the CDW in the $a,b$ plane is then impeded by the changed d--band occupation, but the 1x1x2 spanning vector is still possible. A further argument in favour of the chiral CDW may be the reduced dimensionality of the chiral CDW having only the $P2$ space group, while the ordinary CDW has $P\bar{3}c1$ \cite{Wezel_chirality_and_orbital}. Due to the corkscrew symmetry of the axial propagation vector the electronic structure in the chiral phase lacks a mirror symmetry to the $a,b$ plane. Since the 1x1x2 folding of states implies just this mirror symmetry, the disappearance of the chiral CDW may be mandatory. The 1x1x2 reconstructed phase represents a new phase that has not been described before. Further theoretical and experimental investigations may therefore be worthwhile. 

Optical polarimetry measurements of this modified phase in the surface region could help to clarify this. We also propose to study this effect further by means of electrostatic charge doping \cite{Biscaras_electrostatic_doping_MoS2_2015}, as this would enable a fine control over conduction band population and simultaneously allow for resistivity measurements. In this manner, a charge-density wave based field effect transistor in TiSe$_2$ might be realised.

\section{Summary and Conclusion}

In summary, we have shown that the CDW in TiSe$_2$ visible to ARPES is initially not of 2x2x2, but of 2x2x1 character. The different dimensionalities of CDW and PLD can be explained by anisotropic electron-phonon coupling. These findings suggest a modification to the models \cite{Wezel_chirality_and_orbital, Zenker_Chiral_charge_order_TiSe2_2013} of the chiral CDW in TiSe$_2$. A component in \textit{c}-direction can only be invoked by the introduction of additional charge carriers, at the expense of the in-plane components, which largely disappear. The emergence of this new CDW component can be understood in the view of a mainly exciton-driven CDW, as the energetically more favourable channels for exciton formation, leading to an in-plane CDW, are now blocked. Water adsorption with its associated experimental challenges could be replaced by electrostatic charge doping methods \cite{Biscaras_electrostatic_doping_MoS2_2015}, allowing for a finer control of the conduction band population.

\begin{acknowledgments}
MM May acknowledges support by the fellowship programme of the German National Academy of Sciences Leopoldina under grant no. LPDS 2015-09. Part of the work has been conducted at BESSY II. We thank the staff for technical assistance.
\end{acknowledgments}
%\bibliographystyle{unsrt}
%\bibliography{bibtex-database}

\end{document}